\documentclass[12pt]{iopart}
\usepackage[pdftex]{graphicx}
\usepackage{amssymb}   

\begin{document}

\title{Three-dimensional Finite Difference-Time Domain Solution of Dirac Equation}

\author{Neven Simicevic \footnote[3]{Correspondence should be addressed to Louisiana Tech University, 
PO Box 10348, Ruston, LA 71272, Tel: +1.318.257.3591, Fax: +1.318.257.4228, 
E-mail: neven@phys.latech.edu}}

\address{\ Center for Applied Physics Studies, Louisiana Tech University,
 Ruston, LA 71272, USA}

\begin{abstract}
The Dirac equation is solved using three-dimensional
Finite Difference-Time Domain (FDTD) method. $Zitterbewegung$ and the dynamics of a well-localized
electron are used as examples of FDTD application to the case of free electrons.

\end{abstract}


\maketitle


In this letter, for the first time, full three-dimensional
FDTD method to solve the Dirac equation is described. 
The Finite Difference-Time Domain (FDTD) method is a fast growing
numerical method originally introduced by
Kane Yee \cite{Yee66} to solve Maxwell's equations. In the last two decades the method
was intensively developed primarily in the field of electrodynamics
\cite{Sad92,Kunz93,Sull00,Taf00}, but was, at a much lower scale, also extended to other fields
of applications such as acoustics and elastodynamics.
When applied to solving  Maxwell's equations, the FDTD method is relatively simple
and numerically robust, has almost no limit in the description of
geometrical and dispersive properties of the material,
and is appropriate for the computer technology of today.
As an example, we have applied the FDTD method to calculate the exposure of
complex biological tissues to non-ionizing
ultra-wide band (UWB) radiation using high-resolution description of the geometry and
realistic physical properties of exposed material over a broad frequency
range \cite{Sim05,Sim08}. 

In the case of electrodynamics,
the FDTD method was able to solve problems with complexity that was far beyond
allowing analytical solutions and was fundamental to the advancement
of electrical engineering \cite{Taf00}.
In the same sense, we expect that the application of FDTD method in quantum mechanics,
in this particular case for the solution of the Dirac equation, will become a stepping stone
for the advancement of modern physics.

Similarities between Maxwell's equations and the Dirac equation are obvious if
the Dirac equation is written, in a standard notation, as a system of two first-order equations \cite{Sak87}
\begin{eqnarray}
{[\imath \hbar {\bf{\sigma \cdot \nabla}} - \imath {\frac{\hbar}{c}} {\frac{\partial}{\partial t}}] \Phi^{(L)}
= -mc \Phi^{(R)}},
\nonumber\\
{[-\imath \hbar {\bf{\sigma \cdot \nabla}} - \imath {\frac{\hbar}{c}} {\frac{\partial}{\partial t}}] \Phi^{(R)}
= -mc \Phi^{(L)}}.
\label{Dir_2comp}
\end{eqnarray}
Two component wave functions $\Phi^{(R)}$ and $\Phi^{(L)}$ couple in these equations as
the electric $\vec E$ and magnetic $\vec B$ fields couple in Maxwell's equations \cite{Sak87}.

While the FDTD scheme can be applied to Eq. (\ref{Dir_2comp}),
it is applied here to the form originally written by Dirac. In the case when the
electromagnetic field described by the four-potential ${A^{\mu}=\{A_{0}(x),\vec A(x)\}}$
is minimally coupled to the particle, the Dirac equation can be written as
\begin{equation}
{\imath \hbar {\frac{\partial \Psi}{\partial t}}= ({H}_{free}+{H}_{int}) \Psi},
\label{Dirac_eq}
\end{equation}
where
\begin{equation}
{{H}_{free} = -\imath c\hbar {{\bf\alpha} \cdot \nabla} + \beta m c^{2}},
\end{equation}
\begin{equation}
{{H}_{int} = - e {{\bf\alpha} \cdot {\vec A} } + e A_{0}},
\end{equation}
and
\begin{equation}
{\Psi (x) =\left( \begin{array} {c} \Psi_{1} (x) \\ \Psi_{2} (x)
\\ \Psi_{3} (x)\\ \Psi_{4} (x) \end{array} \right)}.
\end{equation}
Matrices {${\bf\alpha}$ and $\beta$ are expressed using $2 \times 2$ Pauli
matrices $\bf\sigma^{'}s$ and the $2 \times 2$ unit matrix $I$ \cite{Sak87}.

The FDTD schematics to solve
Eq. (\ref{Dirac_eq}) follows Yee's leapfrog algorithm \cite{Yee66}.
Space and time are discretized using uniform rectangular lattices of size
$\Delta x$, $\Delta y$ and $\Delta z$, and uniform time increment $\Delta t$.
Any function $f(x,y,z,t)$ becomes
$f(i\Delta x,j\Delta y,k\Delta z,n \Delta t) \equiv f^{n}_{i,j,k}$, where i,j,k, and n are integers.
Partial derivatives are expressed numerically with second order accuracy as follows:
\begin{itemize}
\item space derivative in the x-direction at fixed time $n \Delta t$
\begin{equation}
{\frac{\partial f^{n}_{i,j,k}}{\partial x}} \approx {\frac{f^{n}_{i+1,j,k}-f^{n}_{i-1,j,k}}
{2\Delta x}}.
\label{Diff_x}
\end{equation}
Analogously for derivatives in y- and z-directions.
\item time derivative at fixed position  $f(i\Delta x,j\Delta y,k\Delta z)$
\begin{equation}
{\frac{\partial  f^{n}_{i,j,k}}{\partial t}} \approx {\frac{{f^{n+1/2}_{i,j,k}-f^{n-1/2}_{i,j,k}}} {\Delta t}}
\label{Diff_t}
\end{equation}
\end{itemize}
In the case of electrodynamics the electric field $\vec E$ at the time $n-1/2$ is used to calculate magnetic field $\vec H$ at the time $n$, which in turn is used to calculate
the electric field $\vec E$ at the time $n+1/2$, and so on. In the case of the Dirac equation the
wave functions $\Psi_{1}$ and $\Psi_{2}$ at the time $n-1/2$ are used to calculate the
wave functions $\Psi_{3}$ and $\Psi_{4}$ at the time $n$, which are then used to calculate
the wave functions $\Psi_{1}$ and $\Psi_{2}$ at the time $n+1/2$, and so on. The same numerical requirements as in the case of electrodynamics were followed. Numerical dispersion
and stability criteria developed for the electromagnetic case \cite{Taf00} were analogously applied to solving the Dirac equation. The Courant stability
condition was imposed to the time step $\Delta t$
\begin{equation}
\Delta t \leq {\frac{1}{c \sqrt{(\Delta x)^{-2}+(\Delta y)^{-2}+(\Delta z)^{-2}}}},
\label{Cond1_t}
\end{equation}
and, as a consequence of the Nyquist sampling limit \cite{Kunz93}, the rectangular lattices size
\begin{equation}
\Delta x \sim \Delta y \sim \Delta z < {\lambda/2}.
\label{Cond1_x}
\end{equation}
$\lambda$ is the wavelength of the plane-wave solution. In practical applications the size of the lattice is typically between $\lambda/10$ and
$\lambda/20$. In the case of $Zitterbewegung$, the spatial oscillation of
the wave packet due to an interference between the positive and negative
energy components \cite{Sak87}, one more condition was imposed to the lattice sizes
\begin{equation}
\Delta x \sim \Delta y \sim \Delta z < { \frac{\hbar}{2mc}}.
\end{equation}

Boundary conditions appropriate for extending the finite computational domain to
infinity are applied by extrapolations similar to Liao extrapolation \cite{Liao84}. The physical
solutions inside the computational domain are extended outside the domain using
Newton's interpolation polynomials \cite{Taf00}.

The updating difference equation for $\Psi_{1}$ was obtained by solving the
algebra of Eq. (\ref{Dirac_eq}), applying Eq. (\ref{Diff_x}) and (\ref{Diff_t}),
and simplifying using $\Delta x = \Delta y = \Delta z$

\begin{eqnarray}
 \Psi_{1}^{n+1/2}(I,J,K)&=&\frac{2-C^{n}(I,J,K)}{C^{n}(I,J,K)}\Psi_{1}^{n-1/2}(I,J,K) \nonumber \\
&-&{\frac{c\Delta t}{2\Delta x C^{n}(I,J,K)}}
[ \Psi_{3}^{n}(I,J,K+1)-\Psi_{3}^{n}(I,J,K-1) \nonumber \\
&+&\Psi_{4}^{n}(I+1,J,K)-\Psi_{4}^{n}(I-1,J,K)-i(\Psi_{4}^{n}(I,J+1,K) \nonumber \\
&-&\Psi_{4}^{n}(I,J-1,K))]
+i{\frac{e\Delta t}{\hbar C^{n}(I,J,K)}}[ A^{n}_{1}(I,J,K)\Psi_{4}^{n}(I,J,K) \nonumber \\
&-&iA^{n}_{2}(I,J,K)\Psi_{4}^{n}(I,J,K)
+A^{n}_{3}(I,J,K)\Psi_{3}^{n}(I,J,K)],
\end{eqnarray}

where $C^{n}(I,J,K)=1+i\frac{\Delta t}{2\hbar}[mc^{2}+eA^{n}_{0}(I,J,K)]$.
Updating equations for $\Psi_{2}$, $\Psi_{3}$, and $\Psi_{4}$ were
constructed in a similar way. The dynamics of a Dirac electron can be now studied
in an environment described by any four-potential $A^{\mu}$ regardless
of its complexity and time dependency.
As a result of limited space in this letter only few cases were chosen.

Following the analytical solution of K. Huang \cite {Huang52}, we studied the model-independent 
motion of a free Dirac electron acquired from the Dirac
equation itself. It is important to stress that, as a consequence of the
Dirac equation being linear
in $\partial / \partial t$, the entire dynamics of the electron is defined by its
initial wave function only. As in the case of Maxwell's equations, no initial
velocity needs to be specified. The dynamics
of the wave packet studied here was defined by its initial wave function

\begin{equation}
{\Psi (\vec x,0) =N \sqrt{\frac{E+mc^{2}}{2E}}\left( \begin{array} {c} 1 \\ 0
\\ \frac{p_{3}c}{E+mc^{2}}\\ \frac{(p_{1}-ip_{2})c}{E+mc^{2}}\end{array} \right)}
e^{-\frac{\vec x \cdot \vec x }{4x_{0}^{2}}+\frac{i\vec p \cdot \vec x}{\hbar}},
\label{Wave_packet}
\end{equation}
where $N$ is a normalizing constant, $N=[(2\pi)^{3/2}x_{0}^{3}]^{-1/2}$. Eq. (\ref{Wave_packet})
represents a wave packet whose initial probability distribution is of a normalized Gaussian
shape with its size defined by the constant $x_{0}$. Its spin is pointed along the z-axis and its
motion is defined by the values of $p_{1},p_{2}$, and $p_{3}$. In the
"single particle interpretation"
of the Dirac equation, if we choose $p_{2}=p_{3}=0$ and $p_{1} \neq 0$ the wave packet should move
in +x-direction. This is not the case. Because of the localization of the wave packet
and limitation of the direction of the spin, the initial condition in
Eq. (\ref{Wave_packet}) contains also a significant
negative energy component moving in the -x-direction \cite {Huang52}.
Dynamics of this wave packet, for $p_{1}=18.75 \; MeV/c$ and $x_{0}=10^{-11}\; m$,
is shown in Fig. \ref{fig:Wave_motion}.

\begin{figure}
\begin{center}
\vspace*{-2.5cm}
\includegraphics[scale=0.35]{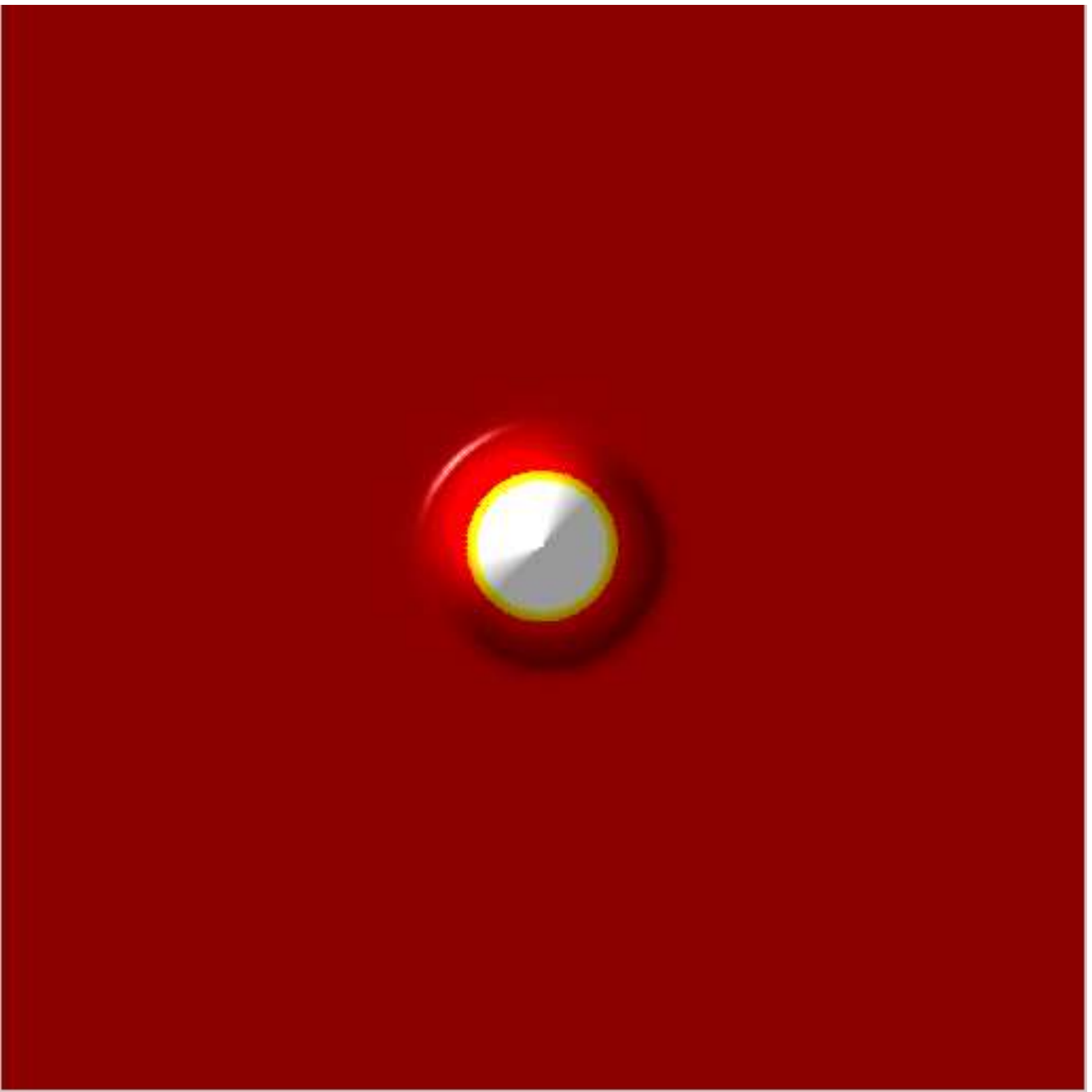}
\includegraphics[scale=0.35]{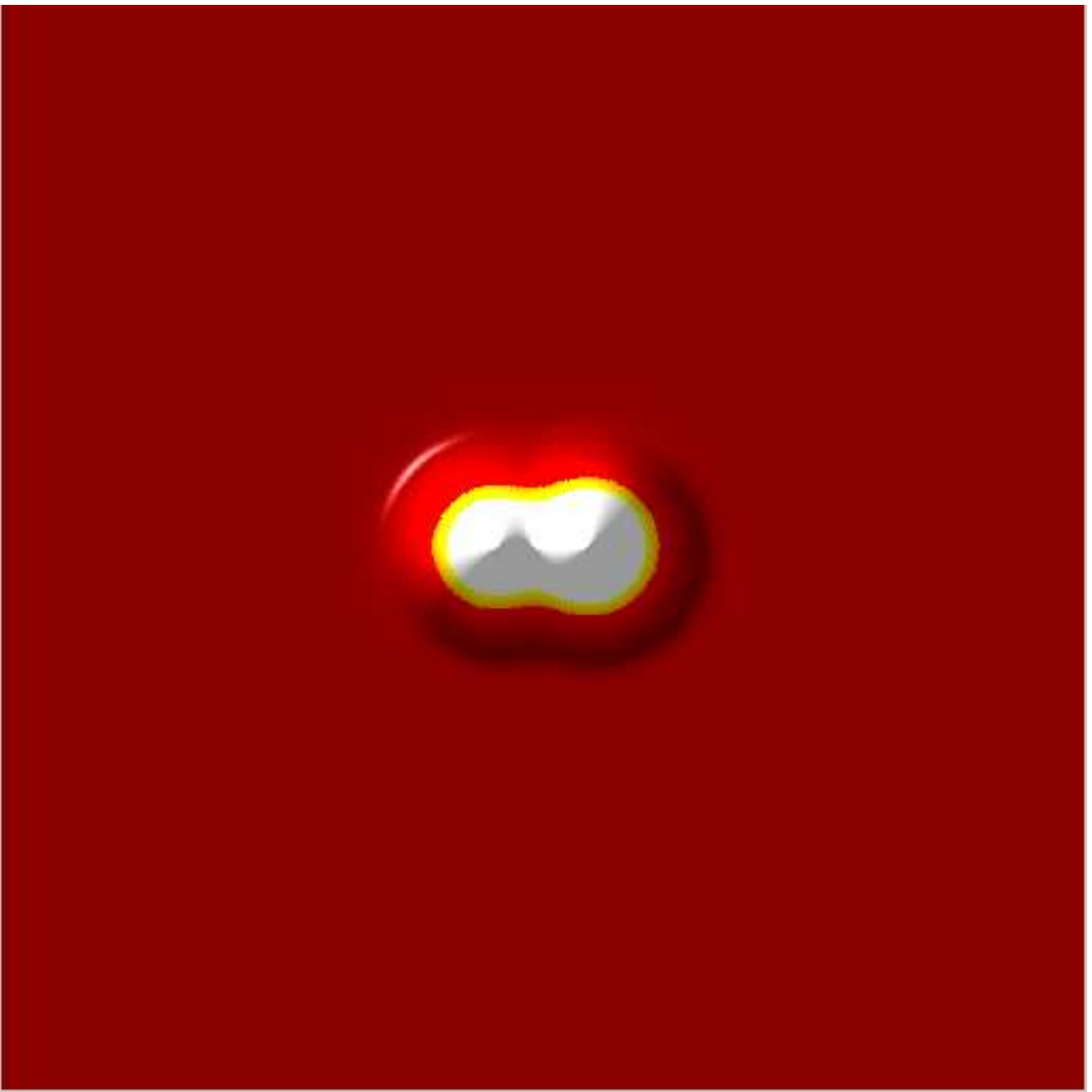}
\end{center}
\vspace*{-3.cm}
\begin{center}
\includegraphics[scale=0.35]{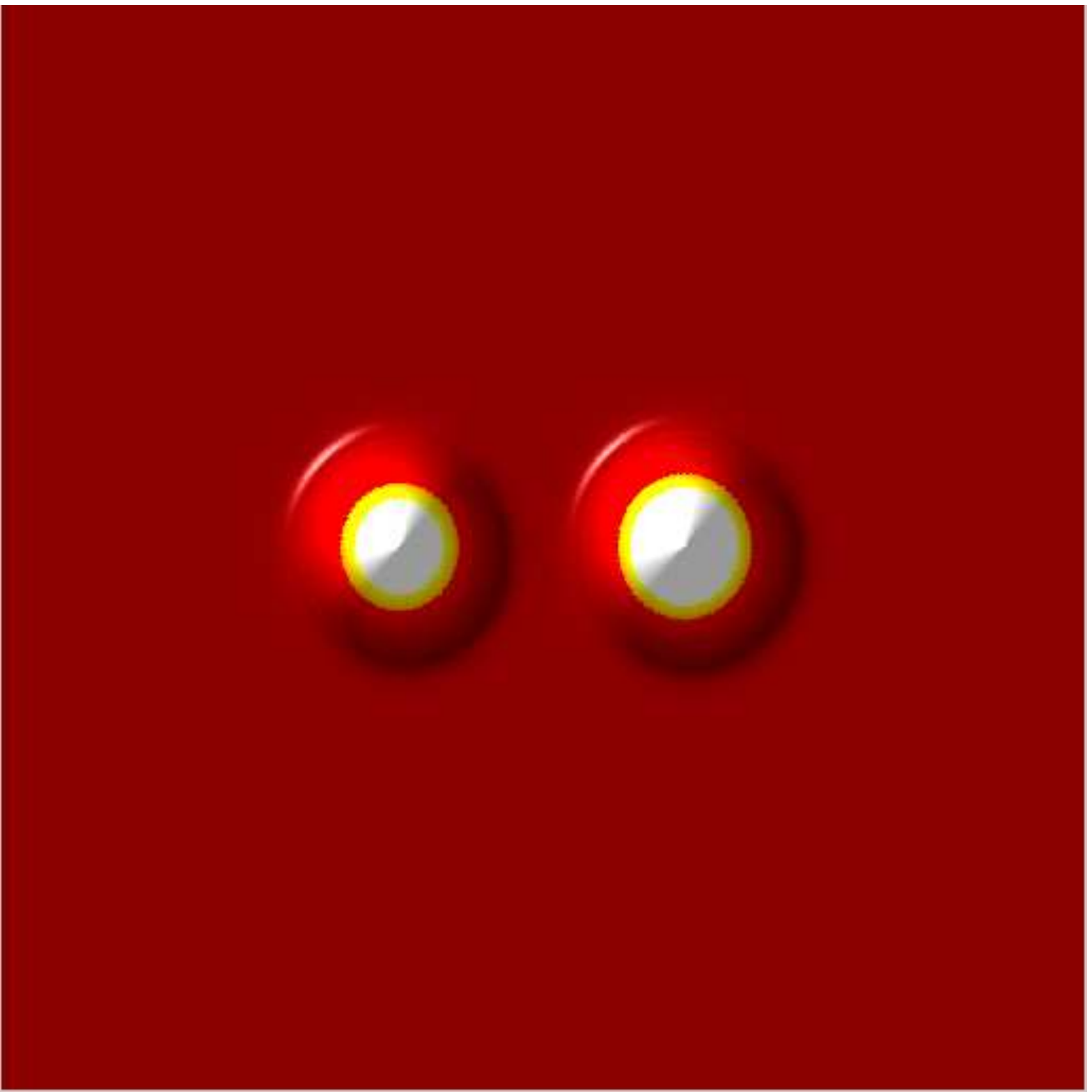}
\includegraphics[scale=0.35]{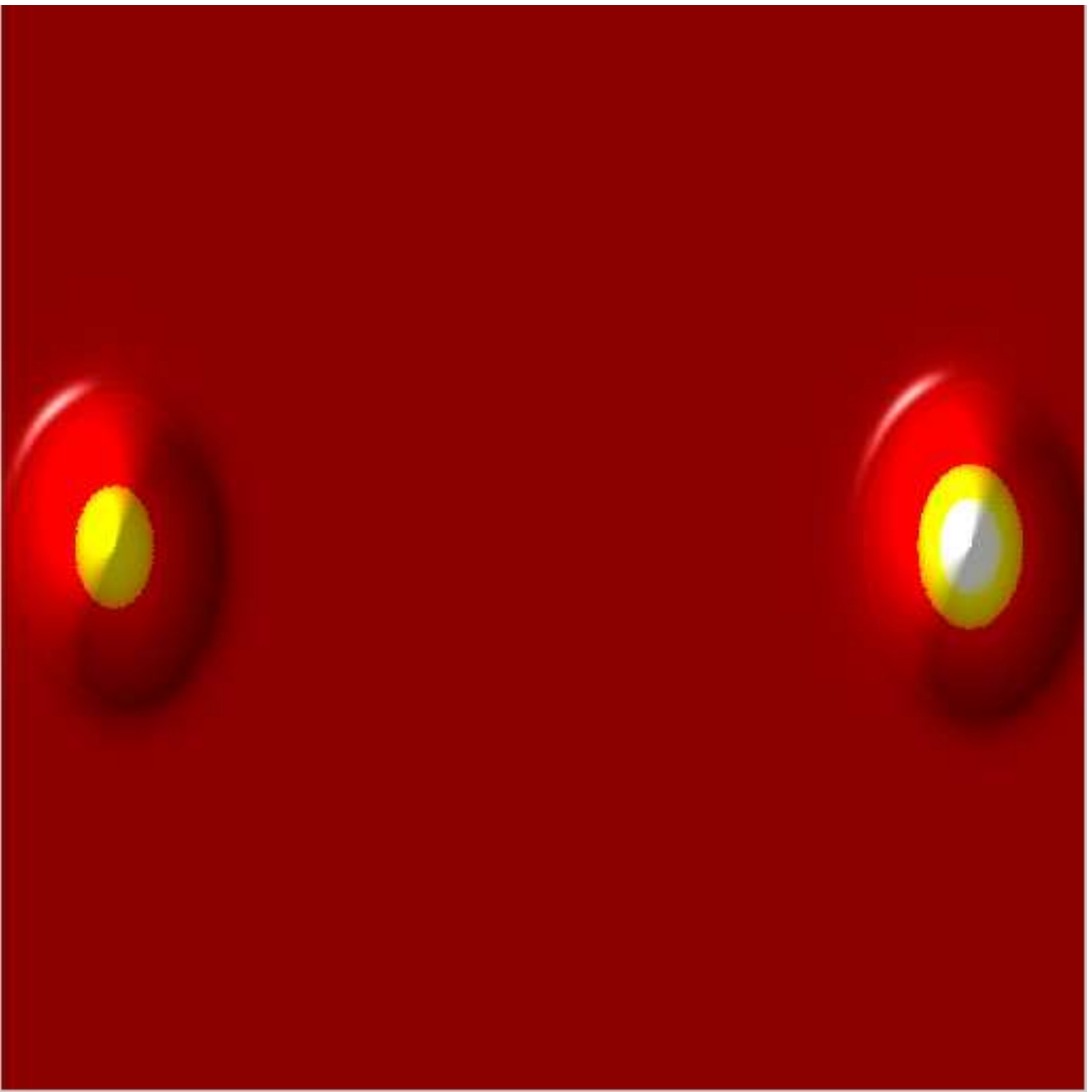}
\end{center}
\vspace*{.5cm}
\caption{Top view of the four different stages of motion of
the wave packet initialized by Eq. (\ref{Wave_packet}) in the x-y plane.
The initial wave packet splits in two, moving in both the positive and negative x-directions,
corresponding to positive and negative energy solutions.  It is important to notice
details of electron dynamics through changing of the shape of the wave packets into
elliptical as a result of relativistic suppression of the wave packet spreading
in the direction of motion for a massive particle \cite{Bloh82}.
The full animations can be accessed on-line \cite{Simi08}.}
\label{fig:Wave_motion}
\end{figure}

Another aspect of the motion of the Dirac electron consists of $Zitterbewegung$,
rapid oscillatory motion around a uniform rectilinear motion attributed to
interference between positive and negative energy states \cite{Sak87}.
The FDTD, being a time-domain method, is an excellent tool to study the properties
of $Zitterbewegung$. Few results are presented here. Fig. \ref{fig:Zitter1} shows
the time dependence of the
position of the center of probability in the x-y plane of the fraction of our wave-packet
moving in the -x-direction at the beginning of motion. The amplitude of the motion
in this particular case is
$1.45 \times 10^{-11} \; cm$, of the order of $\hbar/mc$ as predicted \cite{Sak87}.
Fig. \ref{fig:Zitter2} shows the x-position of the fraction of the wave-packet
moving in the +x-direction.
$Zitterbewegung$ exists at the beginning of the motion, at the time
when positive- and negative-energy components of the wave packets shown in
Fig. \ref{fig:Wave_motion} overlap. It is absent as the wave packets separate.
Lack of space in this letter prevents discussion of  $Zitterbewegung$-related violent
and rapid oscillations of the velocity of the electron.

\begin{figure}
\begin{center}
\vspace*{-3.5cm}
{\scalebox{.5}{\includegraphics{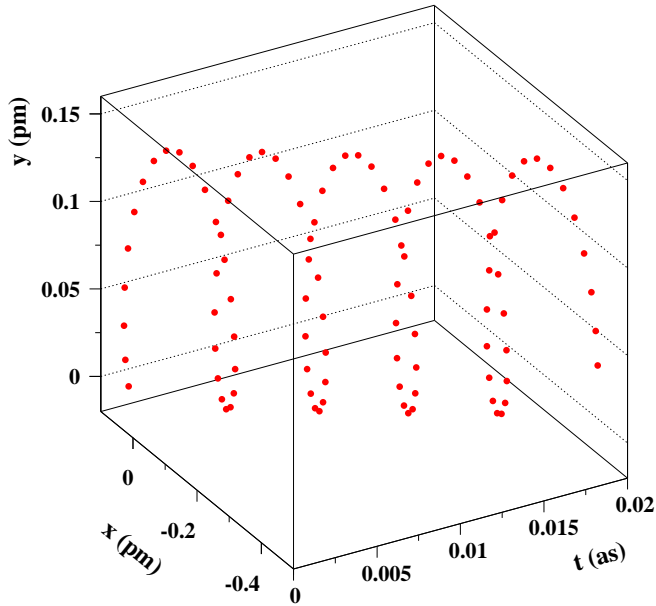}}}
\end{center}
\caption{\label{fig:Zitter1} Time dependence of the position of the center of
probability of the fraction of the wave-packet defined by Eq. (\ref{Wave_packet})
moving in the -x-direction at the beginning of motion.}
\end{figure}

\begin{figure}
\begin{center}
\vspace*{-3.5cm}
{\scalebox{.5}{\includegraphics{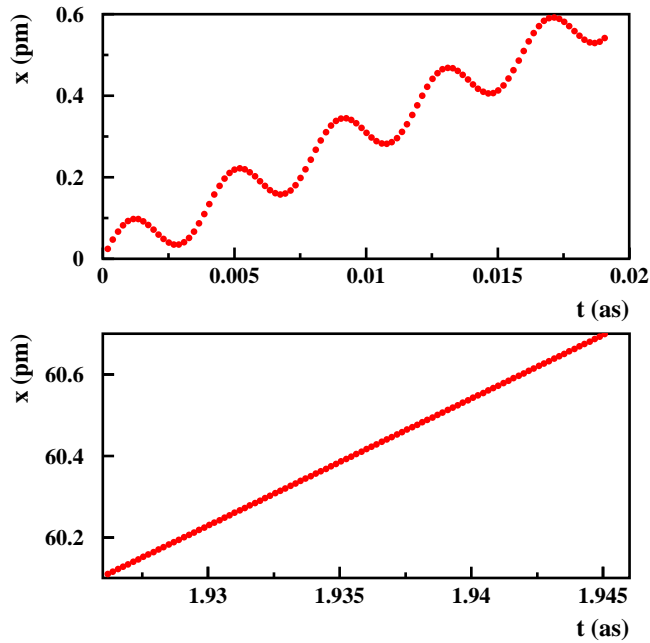}}}
\end{center}
\vspace*{-1.cm}
\caption{\label{fig:Zitter2} X-position of the fraction of the wave-packet shown
in Fig. \ref{fig:Wave_motion} moving in the +x-direction.
The $Zitterbewegung$ which exists at the beginning of the motion (top)
is absent after positive- and negative-energy components separate (bottom).}
\end{figure}

The frequency
of $Zitterbewegung$ can easily be obtained.
Fitting the probability function, as shown in Fig. \ref{fig:Zitter3}, we found
the angular frequency of $\omega=(1.582 \pm 0.009) \times 10^{21} \; Hz$
for $p_{1}=18.75 \; MeV/c$ and
$\omega=(1.525 \pm 0.009) \times 10^{21} \; Hz$  for $p_{1}=187.5 \; eV/c$.
It is in agreement with expected angular frequency of $\sim 2mc^{2}/\hbar$ and
shows little momentum dependence.

\begin{figure}
\begin{center}
\vspace*{-5.0cm}
{\scalebox{.5}{\includegraphics{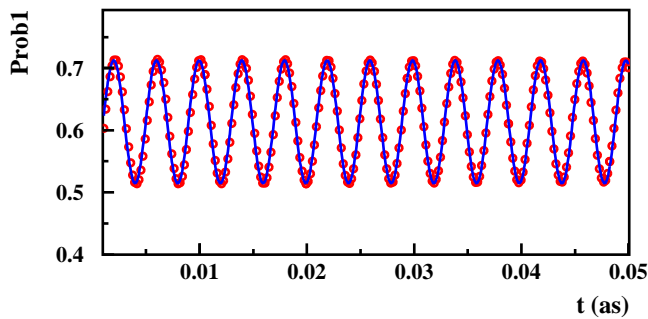}}}
\end{center}
\vspace*{- 5.cm}
\caption{\label{fig:Zitter3} Fit of the probability function used to determine
the frequency of $Zitterbewegung$.}
\end{figure}

Finally, we discuss briefly the well-localized state
described, for example, by
\begin{equation}
{\Psi (\vec x,0) =N \left( \begin{array} {c} 1 \\ 0
\\ 0 \\ 0 \end{array} \right)}
e^{-\frac{\vec x \cdot \vec x }{4x_{0}^{2}}}.
\label{Wave_packet2}
\end{equation}
Snapshots of the dynamics of the probability densities of this state for $x_{0}=10^{-14}\; m$,
simulating localization in a volume of the size of the atomic nucleus, are shown in
Fig. \ref{fig:Spherical_wave}. FDTD computation shows that the angular distribution of
probability density $|\Psi_{1}|^2$ is determined by spherical harmonic $Y_{00}$,
the density $|\Psi_{2}|^2$ is 0, the angular distribution of $|\Psi_{3}|^2$
is determined by spherical harmonic $Y_{10}$, and
$|\Psi_{4}|^2$ by $Y_{11}$. The radial dependence of $|\Psi_{1}|^2$ is determined by
the spherical Bessel functions $j_{0}$, and $|\Psi_{3}|^2$ and $|\Psi_{4}|^2$ by $j_{1}$.
The time dependent behavior of the total probabilities is shown in
Fig. \ref{fig:Spherical_wave_prob}. After the first shock,
attributed to $Zitterbewegung$ resulting from the initial condition,
total probabilities acquire the values of $Prob1 : Prob2 : Prob3 : Prob4 = 1/2 : 0 : 1/6 : 1/3$.
The ratios of those probabilities result from the  Clebsch-Gordon coefficients which
are part of the relation between spherical harmonic spinors $\Omega_{jlm}$
and spherical harmonics $Y_{lm}$ \cite{Akhi65,Bere82}.(Those ratios
persist for more localized states, for example for $x_{0}=10^{-15}\; m$, but, as expected, change
for less localized states.)
The angular and radial distribution of the probability densities and the values
of the total probability determined by the FDTD method show that the state with
initially localized position and spin, corrects its state in a very short time
(Fig. \ref{fig:Spherical_wave_prob}) and behaves as a spherical wave.

\begin{figure}
\centering
{\scalebox{.18}{\includegraphics{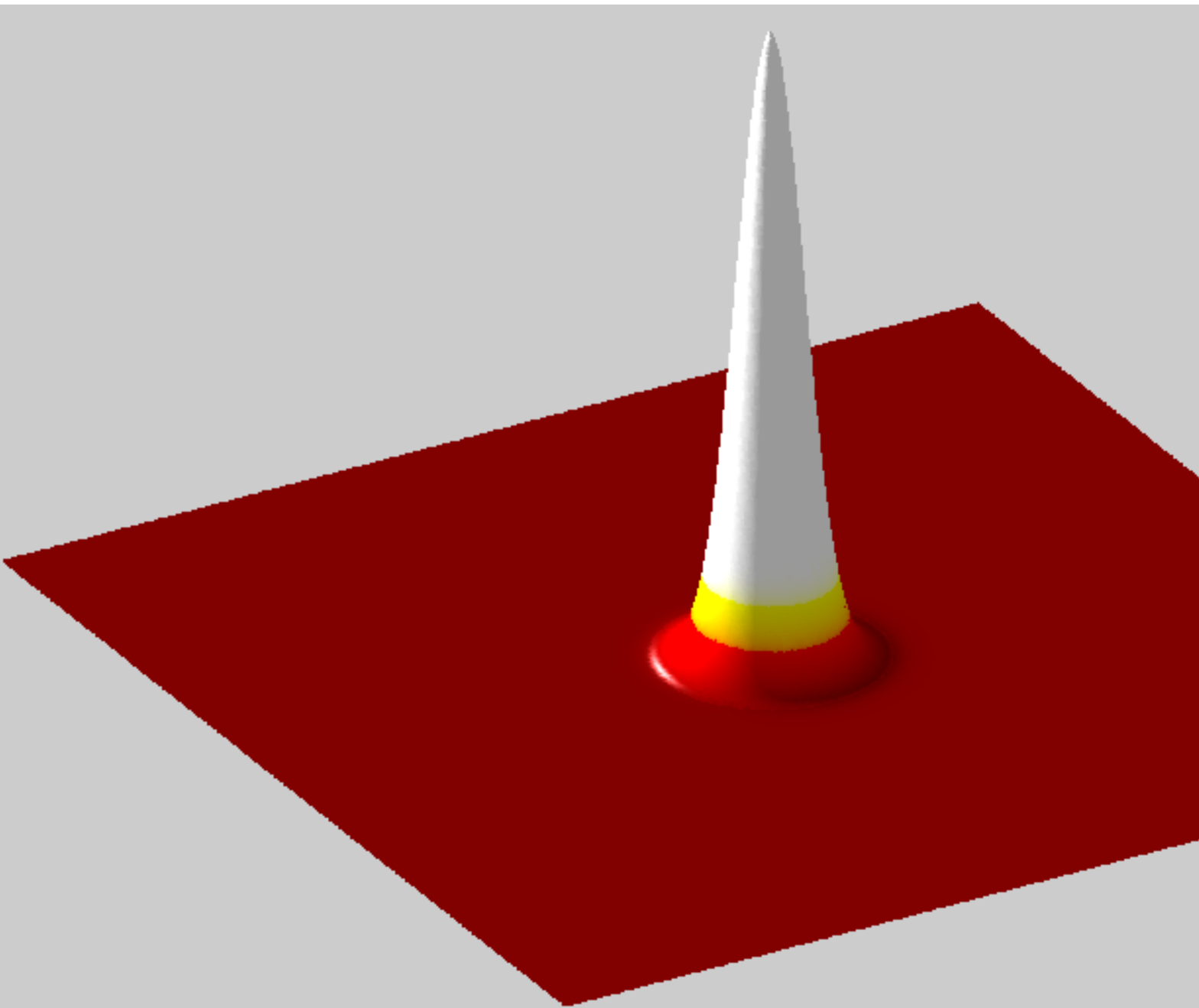}}}
{\scalebox{.18}{\includegraphics{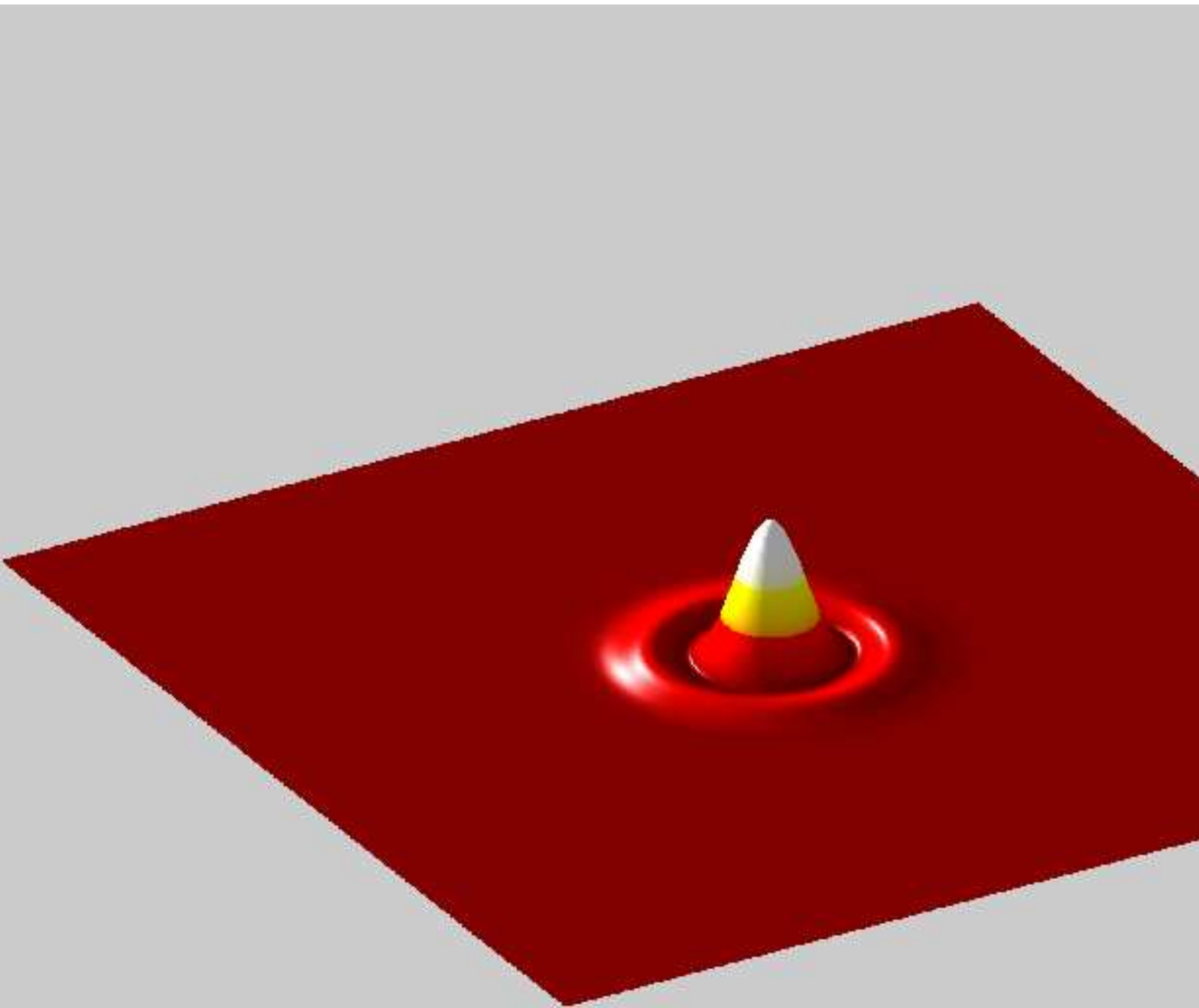}}}
{\scalebox{.18}{\includegraphics{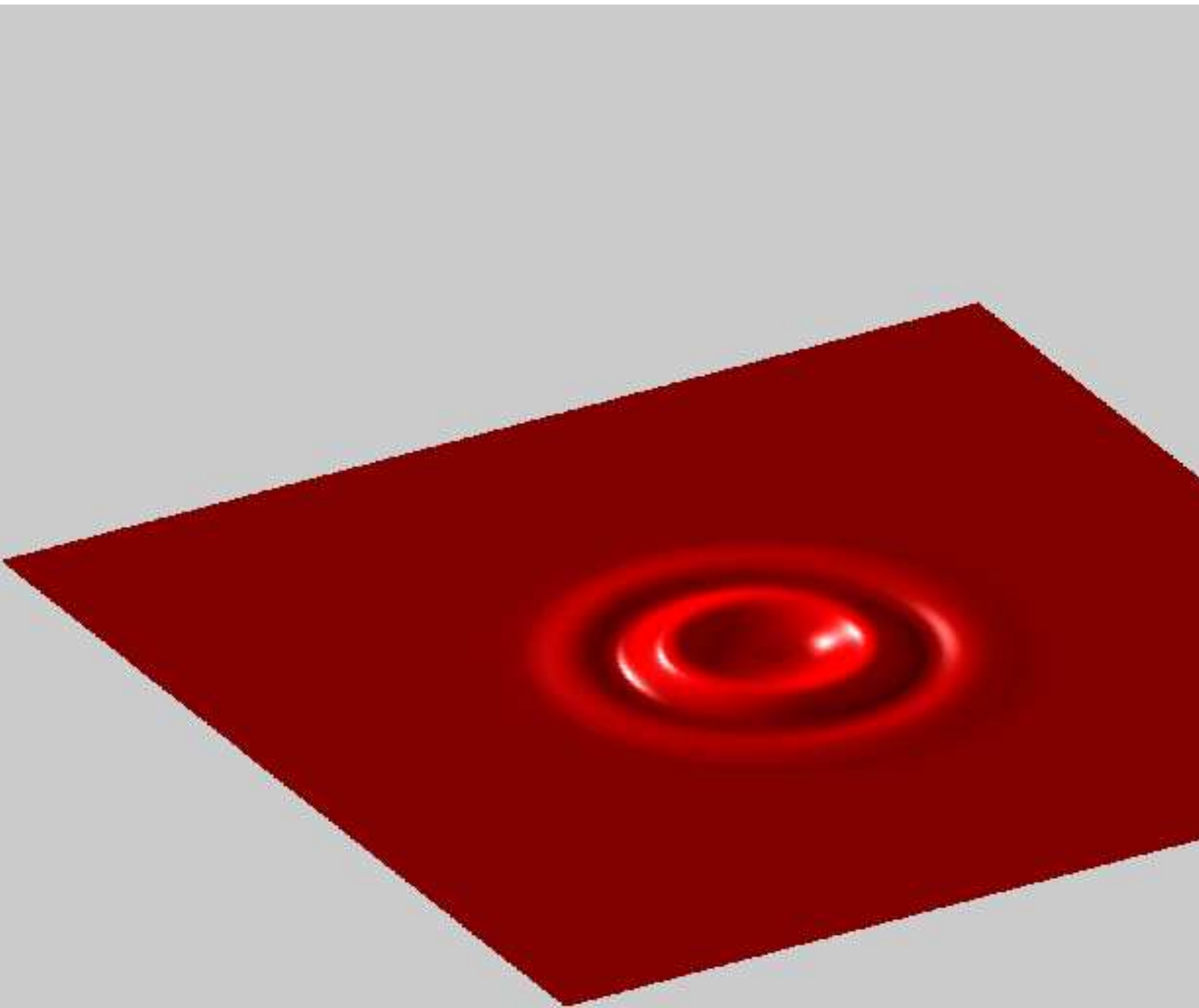}}}
{\scalebox{.18}{\includegraphics{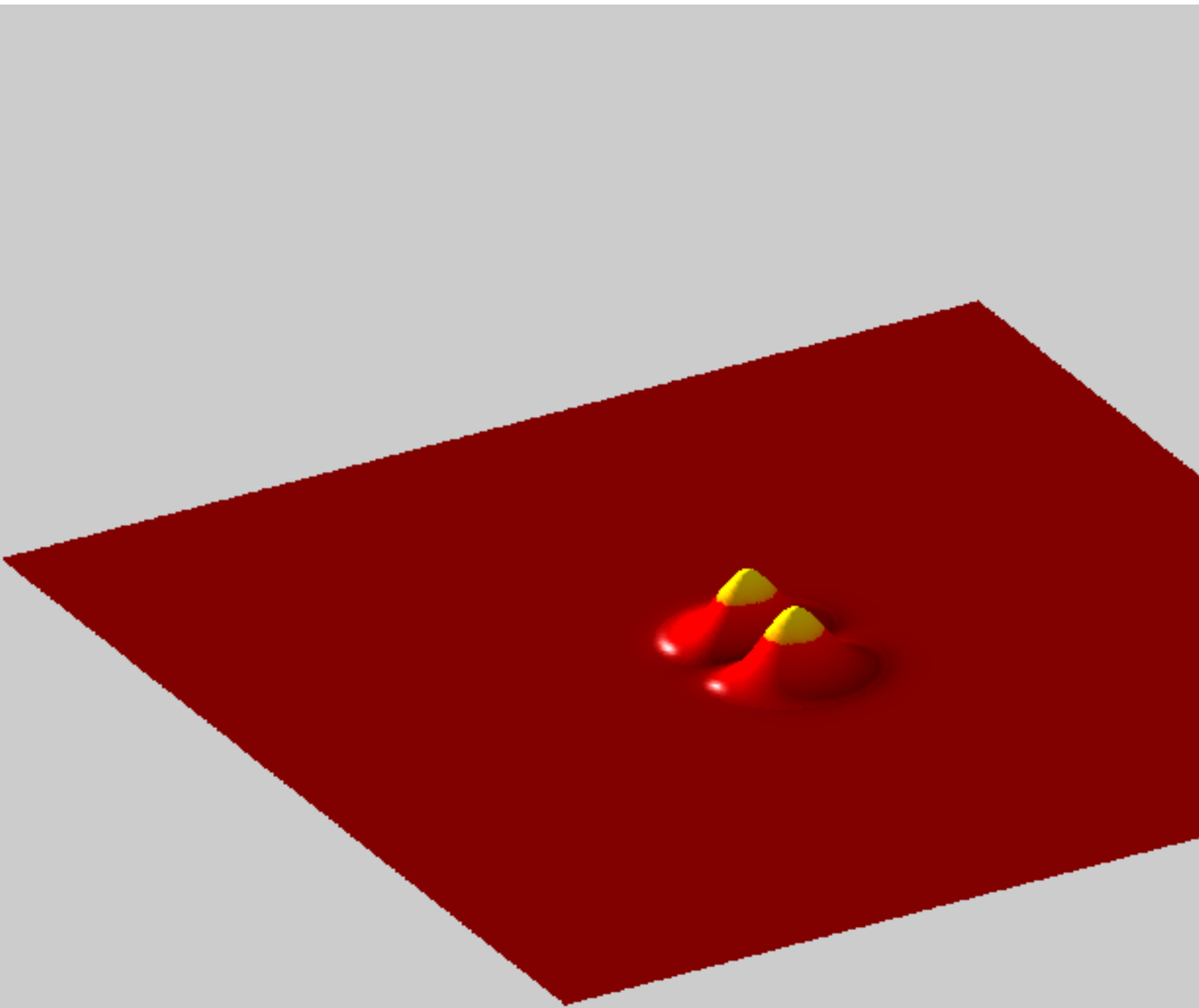}}}
{\scalebox{.18}{\includegraphics{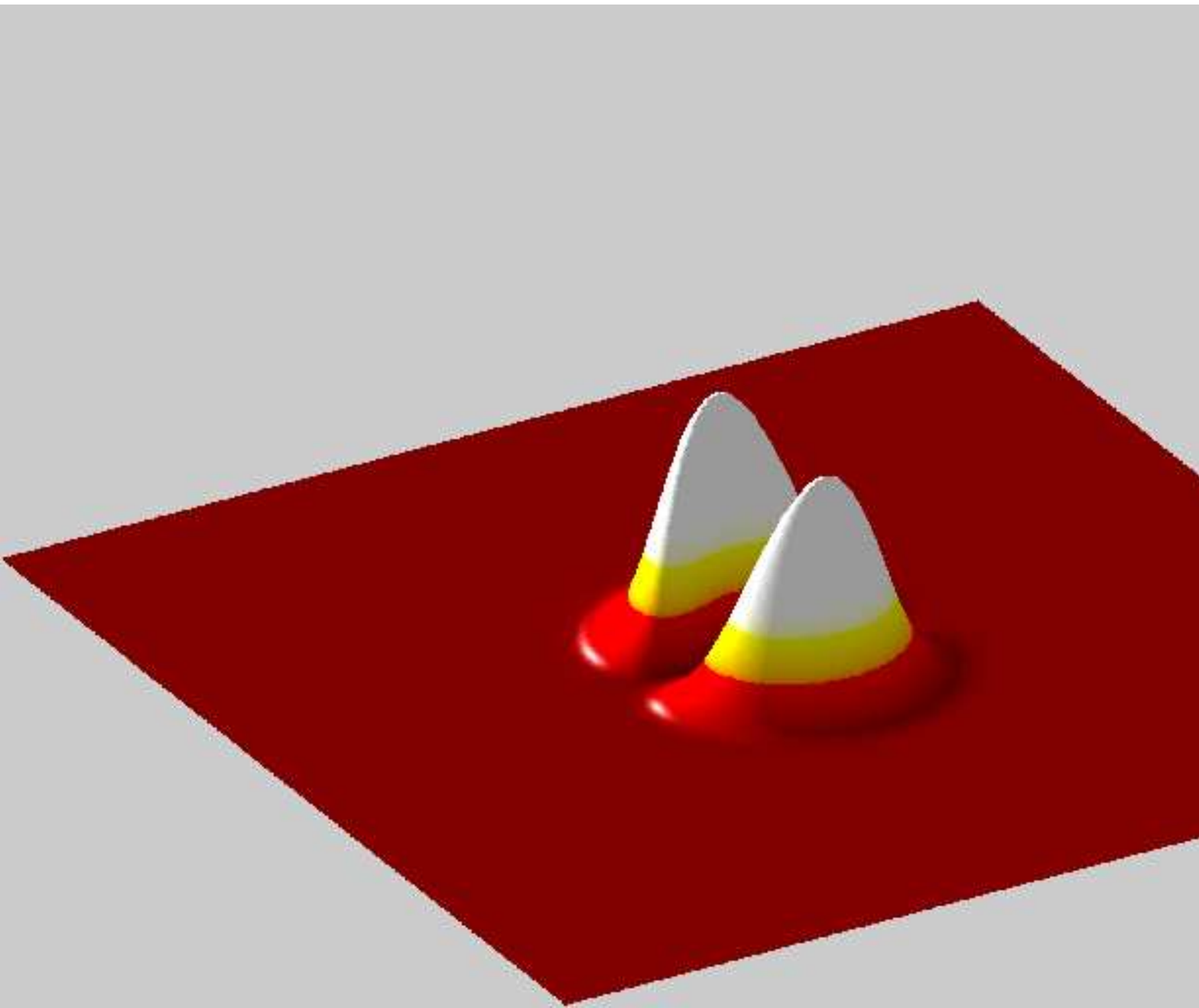}}}
{\scalebox{.18}{\includegraphics{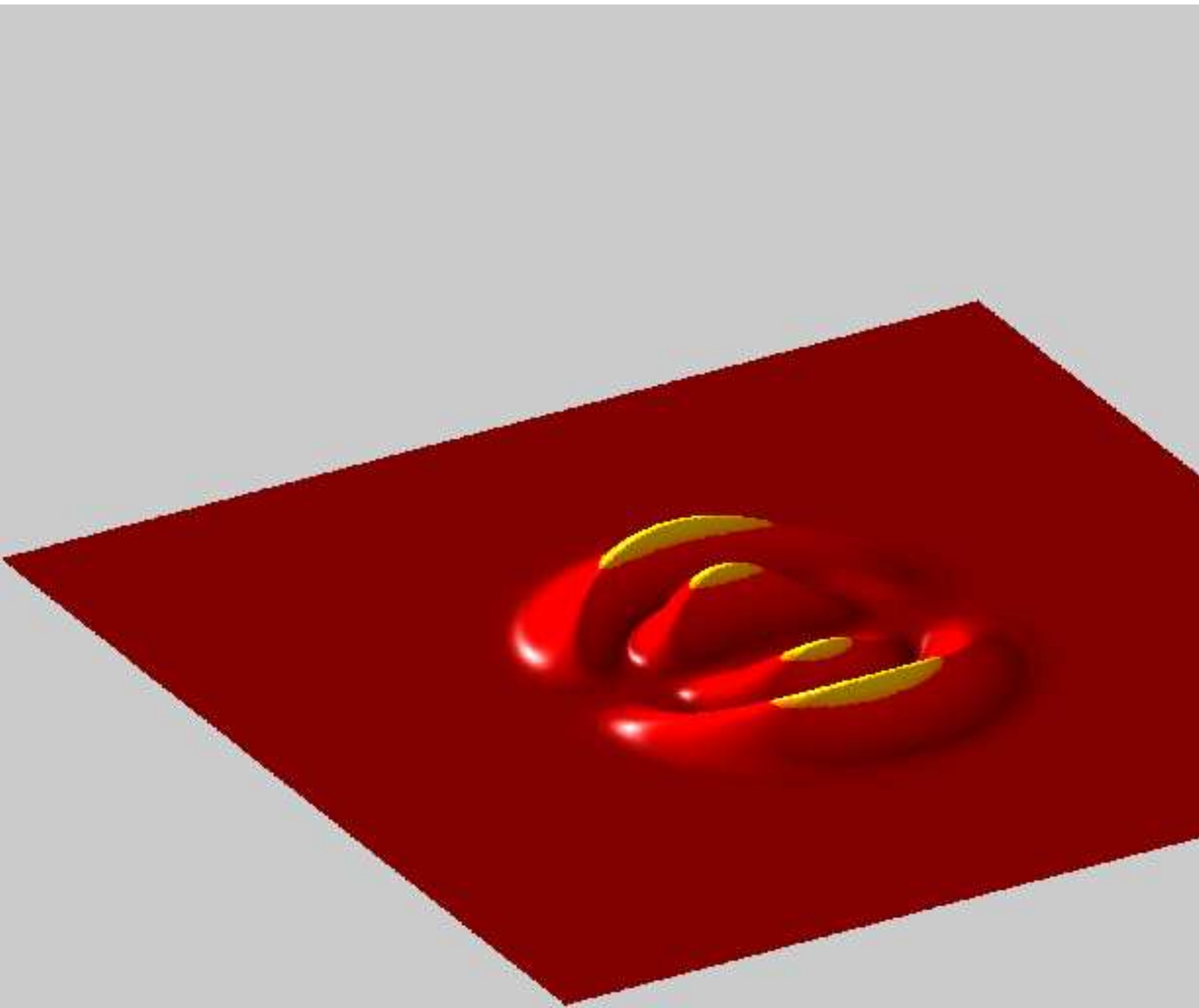}}}
{\scalebox{.18}{\includegraphics{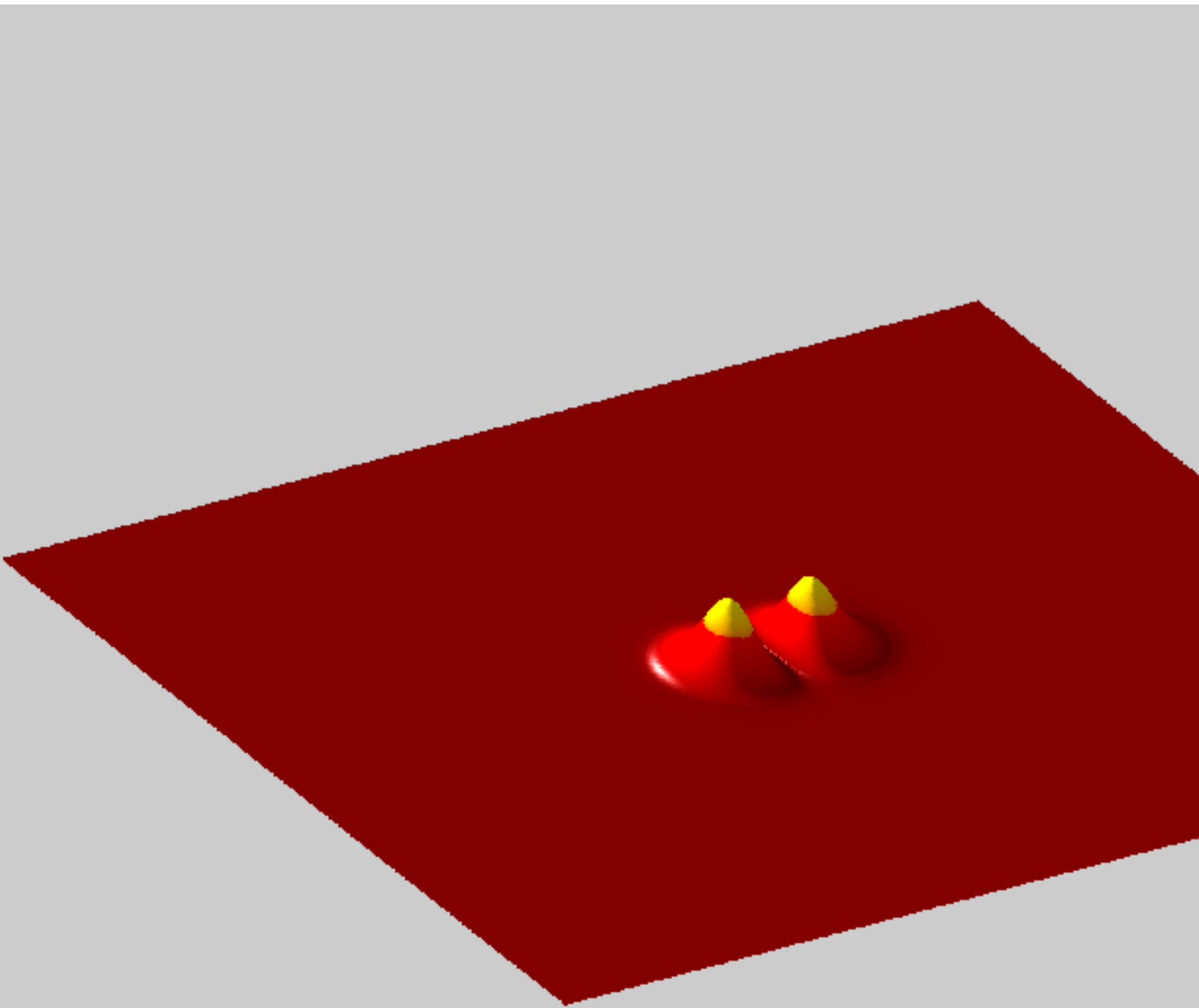}}}
{\scalebox{.18}{\includegraphics{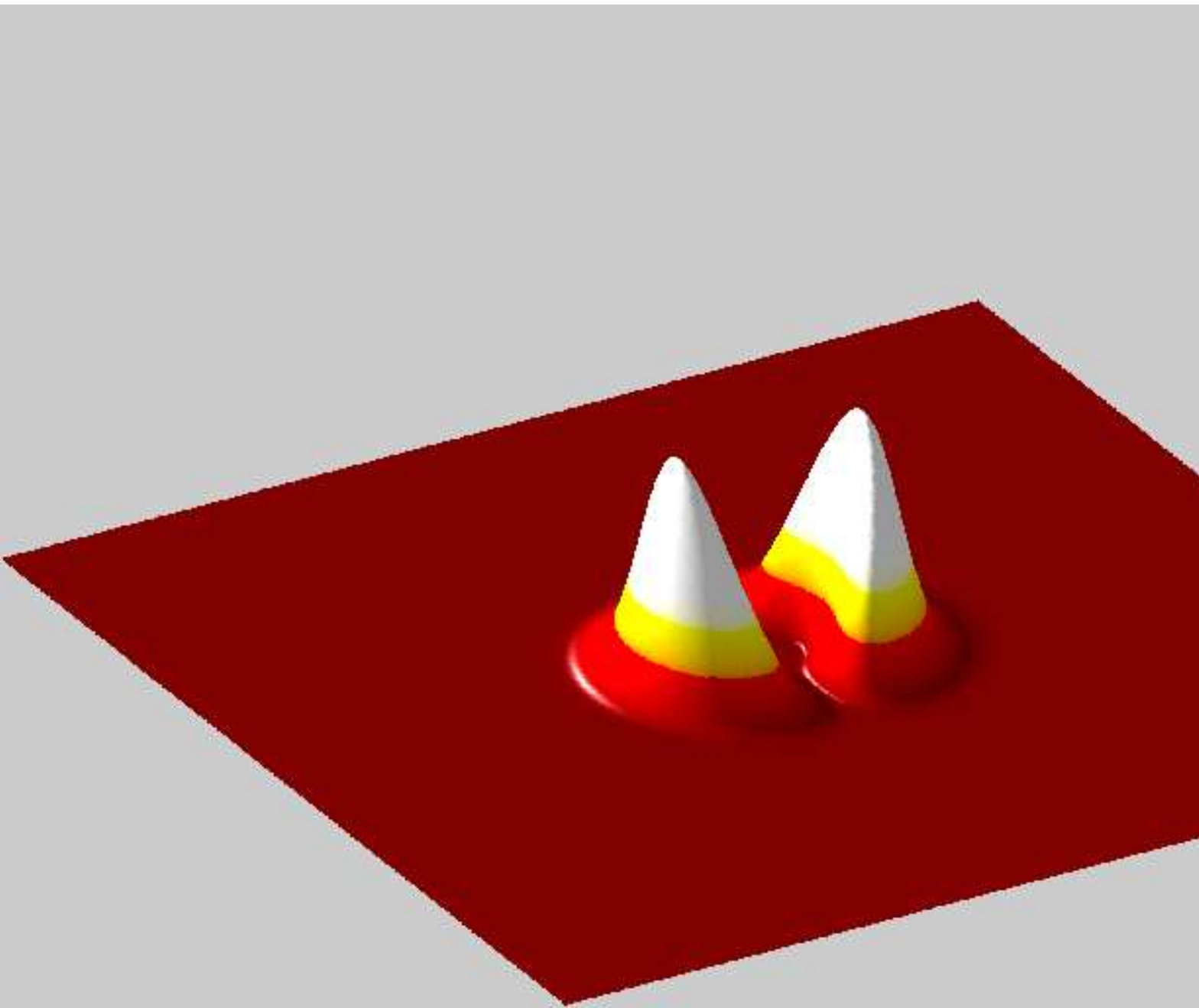}}}
{\scalebox{.18}{\includegraphics{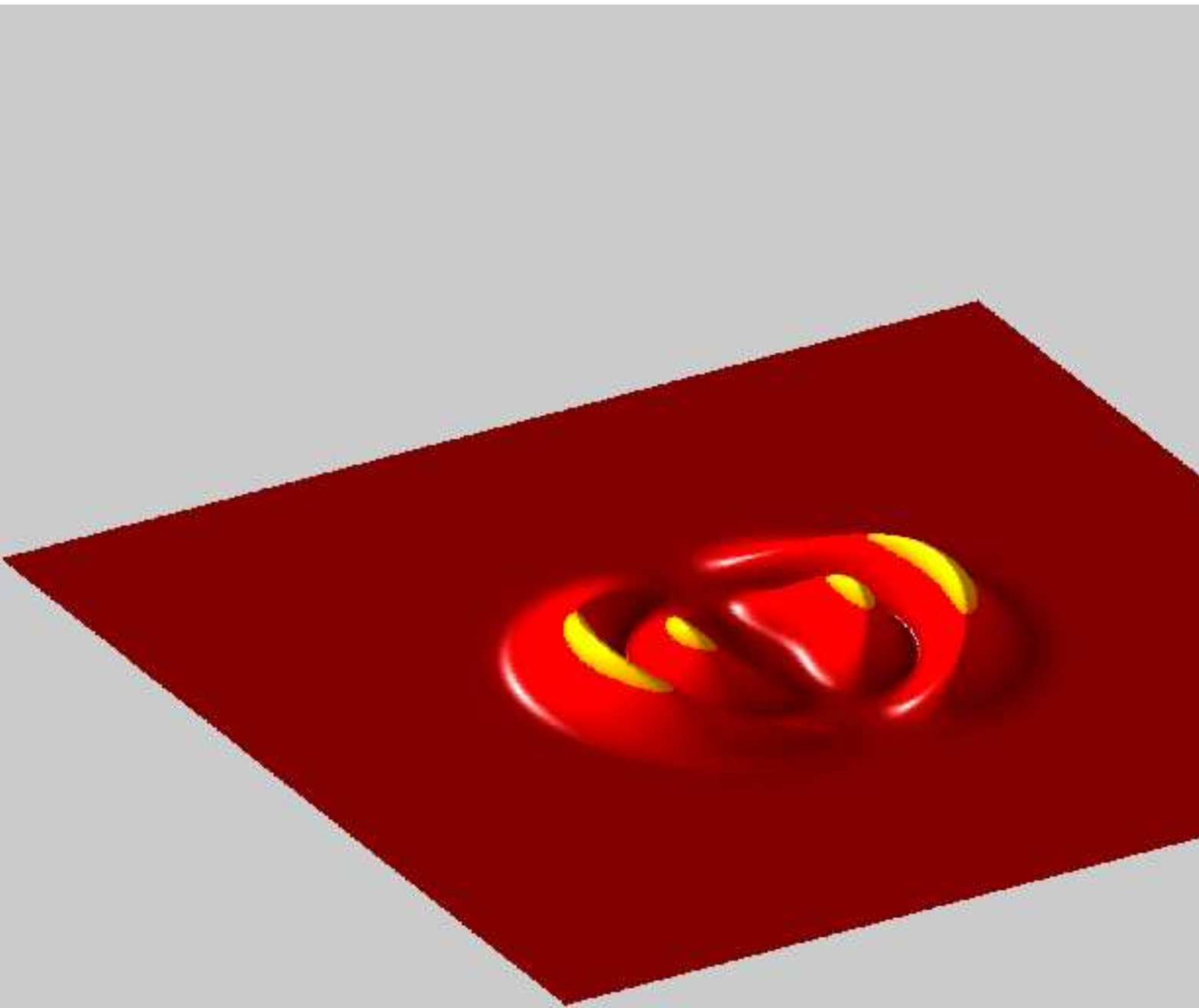}}}
\vspace*{.5cm}
\caption{\label{fig:Spherical_wave} Three different stages of motion of
the wave packet in the x-z plane initialized by Eq. (\ref{Wave_packet2}).
Top-row is probability density $|\Psi_{1}|^2$, mid-row $|\Psi_{3}|^2$,
and bottom-row $|\Psi_{4}|^2$. The associated animations can be accessed on-line \cite{Simi08}.}
\end{figure}

\begin{figure}
\begin{center}
\vspace*{-4.5cm}
{\scalebox{.5}{\includegraphics{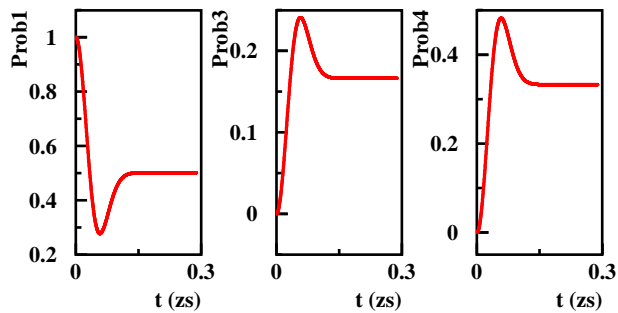}}}
\end{center}
\vspace*{- 4.5cm}
\caption{\label{fig:Spherical_wave_prob} Time dependence of the total probabilities of the wave
packet defined by Eq. (\ref{Wave_packet2}).}
\end{figure}

To conclude, for the first time full three-dimensional Finite Difference Time Domain (FDTD)
method was developed to solve the Dirac equation. In this paper the method was
applied to the dynamics of a free Dirac electron, comparing some of the results to the
analytical solutions. Forthcoming work will study the dynamics of the Dirac electron
in potential fields. The impact of FDTD method in electrodynamics warranted that the first
chapter of the book by A. Taflove and S. C. Hagnes \cite{Taf00} be titled ``Electrodynamics
Entering the 21st Century". We hope that the FDTD method will have the same impact on
better understanding, advancing, and applying modern physics. 

I would like to thank B. Ramu Ramachandran, Lee Sawyer and Steve Wells for useful comments.
Also, the use of the high-performance computing resources provided by Louisiana Optical
Network Initiative (LONI; www.loni.org) is gratefully acknowledged.

\end{document}